# An adapting auditory-motor feedback loop can contribute to generating vocal repetition

Jason Wittenbach[1,#], Kristofer E. Bouchard[2,3,4#], Michael S. Brainard[2,5], Dezhe Z. Jin[1,*]

1. Department of Physics and Center for Neural Engineering, the Pennsylvania State University, University Park, Pennsylvania, USA

2. Department of Physiology and Center for Integrative Neuroscience, University of California at San Francisco, San Francisco, California, USA

3. Department of Neurosurgery and Center for Neural Engineering and Prosthesis, University of California at San Francisco, San Francisco, California, USA.

4. Computational Research Division, Lawrence Berkeley National Laboratory, Berkeley, California, USA

5. Howard Hughes Medical Institute

#: These authors contributed equally to this work.

* Corresponding author: dzj2@psu.edu

**Summary**

Consecutive repetition of actions is common in behavioral sequences. Although integration of sensory feedback with internal motor programs is important for sequence generation, if and how feedback contributes to repetitive actions is poorly understood. Here we study how auditory feedback contributes to generating repetitive syllable sequences in songbirds. We propose that auditory signals provide positive feedback to ongoing motor commands, but this influence decays as feedback weakens from response adaptation during syllable repetitions. Computational models show that this mechanism explains repeat distributions observed in Bengalese finch song. We experimentally confirmed two predictions of this mechanism in Bengalese finches: removal of auditory feedback by deafening reduces syllable repetitions; and neural responses to auditory playback of repeated syllable sequences gradually adapt in sensory-motor nucleus HVC. Together, our results implicate a positive auditory-feedback loop with adaptation in generating repetitive vocalizations, and suggest sensory adaptation is important for feedback control of motor sequences.

**Running title**

Feedback adaptation and repetitive vocalization

**Highlights**

1. Deafening reduces syllable repetition in the Bengalese finch.

2. Auditory responses in sensory-motor area HVC to repeated syllables gradually adapt.

3. Adapting auditory feedback explains the statistics of long syllable repetitions.

**Introduction**

Many complex behaviors – human speech, playing a piano, or birdsong – consist of a set of discrete actions that can be flexibly organized into variable sequences (Doupe and Kuhl, 1999; Lashley, 1951; Rhodes et al., 2004). A feature of many variably sequenced behaviors is the occurrence of repetitive sub-sequences of the same action. Examples include trills in music, repeated syllables in birdsong, and syllable/sound repetitions in stuttered speech. A central issue in understanding how nervous systems generate complex sequences is the role of sensory feedback vs. internal motor programs (Grossberg, 1986) (Fig.1a). At one extreme (the serial chaining framework), the sensory feedback from one action initiates the next action in the sequence; therefore sensory feedback is critical for sequencing the actions (Hull, 1930; James, 1950; Watson, 1925). However, because of the delays in both motor and sensory processing in nervous systems, it has been argued that a sequence generation mechanism relying solely on sensory feedback would be too slow to account for the execution of fast sequences such as typing and speech (Lashley, 1951). At the other extreme, sequences are generated by internal motor programs controlling sequence production without the use of sensory feedback (Grossberg, 1967, 1969; Keele, 1968; Rosenbaum et al., 1983; Summers and Anson, 2009). However, there is ample evidence that sensory feedback can affect action sequences. (Lee, 1950; MacKay, 1986; Pfordresher and Palmer, 2006; Sakata and Brainard, 2006, 2008). Despite the ubiquity of sequencing in behavior, the neural mechanisms of how sensory feedback interacts with internal motor programs to influence discrete actions remain largely unexplored.

Here, we study the role of sensory feedback in the production of repetitive vocal sequences using the Bengalese finch as a model system. The Bengalese finch produces songs composed of discrete acoustic events, termed syllables, organized into variable sequences (Fig.1b). However, sequence production is not random (Okanoya, 2004), as the transition probabilities between syllables are statistically reproducible across time (Sakata and Brainard, 2006; Warren et al., 2012). A prominent feature of the songs of several songbird species, including the Bengalese finch, is syllable repetition (Boughey and Thompson, 1981; Gardner et al., 2005; Jin and Kozhevnikov, 2011; Markowitz et al., 2013; Marler and Peters, 1977; Okanoya, 2004) (e.g. 'b' in Fig.1b). For a given repeated syllable, the number of consecutively produced repeats (the repeat number) varies. The distribution of repeat numbers can violate the predictions of a first order Markov process, with the most probable repeat number (peak repeat number) being much greater than one (Fujimoto et al., 2011; Hampton et al., 2009; Jin and Kozhevnikov, 2011; Markowitz et al., 2013) (Fig. 1c). In the songs of the Bengalese finch, the transition probabilities between syllables are altered shortly after deafening (Okanoya and Yamaguchi, 1997; Woolley and Rubel, 1997) or in real-time by delayed auditory feedback (Sakata and Brainard, 2006), demonstrating that disturbing auditory feedback can disturb sequence generation.

Songbirds are prominent models for studying the neural basis of complex sequence production. Experimental data from sensory-motor song nucleus HVC (proper name) of singing zebra finches have led to neural network models of the internal motor program for sequence generation that instantiate first-order Markov processes (Jin, 2009). This suggests that additional mechanisms contribute to the generation of non-Markovian distributions of repeat numbers (Jin, 2009; Jin and Kozhevnikov, 2011; Markowitz et al., 2013). One possibility is that, because of sensory-motor delays, auditory feedback from the previous syllable interacts with the internal motor program to contribute to the transition dynamics for subsequent syllables (Bouchard and Brainard, 2013; Hanuschkin et al., 2011; Sakata and Brainard, 2006, 2008). For repeated syllables, we hypothesized that the interaction of auditory-feedback and ongoing motor activity forms a positive-feedback loop that contributes to sustaining syllable repetition beyond the predictions of a Markov process (Fig. 1a). However, such positive-feedback architectures are inherently unstable, prone to indefinite repetition (i.e. perseveration). Across sensory modalities, a common feature of sensory responses to repeated presentations of identical physical stimuli is a gradual decrease of response magnitude (i.e. response adaptation). We therefore hypothesized that auditory inputs are subject to response adaptation, which gradually reduces the strength of the positive feedback loop over time. Thus, an auditory-motor feedback loop with response adaptation is predicted to contribute to the generation of non-Makovian repeated syllable sequences by both pushing repeat counts beyond the expectations of a Markov process and simultaneously preventing indefinite repetitions of the syllable. We tested these hypotheses using a combination of neural network and mathematical modeling combined with behavioral and electrophysiological experiments.

# Results

## A network model with adapting auditory feedback

In songbirds, HVC has been proposed to contain an internal motor program for the generation of song sequences (Chang and Jin, 2009; Fee, 2004; Hahnloser et al., 2002; Jin, 2009; Jin et al., 2007; Long and Fee, 2008; Long et al., 2010). HVC sends descending motor commands for song timing to nucleus RA (the robust nucleus of the arcopallium), which in turn projects to brainstem areas controlling the vocal organs (Nottebohm et al., 1976, 1982) (Fig. 2a). HVC also receives input through internal feedback loops from the brainstem (Schmidt, 2003), via Uva (nucleus uvaeformis) and NIf (the interfacial nucleus of the nidopallium) (Lewandowski et al., 2013). Experiments in the zebra finch have shown sparse sequential firing of the RA projecting HVC neurons ($HVC_{RA}$) during singing (Fee, 2004; Hahnloser et al., 2002; Long et al., 2010). This has led to the hypothesis that the motor program for sequence production in HVC includes sequential "chaining" of activity, in which populations of $HVC_{RA}$ neurons responsible for generating a syllable drive the neuronal populations that generate subsequent syllables either directly within HVC or through the internal feedback loop (Abeles, 1991; Chang and Jin, 2009; Fee et al., 2004; Gibb et al., 2009; Long et al., 2010) (Fig 2b).

Our model for generating syllable sequences starts with such a synaptic chain framework. The details of this model have been described previously (Jin, 2009) and are summarized in Materials and Methods. In synaptic chain models, each syllable is encoded in a chain network of $HVC_{RA}$ neurons (Fig. 2b). Spike propagation through the chain produces the encoded syllable by driving appropriate RA neurons. To generate variable syllable transitions, the syllable-chains are connected into branching patterns. At a branch point, syllable-chains compete with each other through a winner-take all mechanism mediated by the inhibitory HVC interneurons ($HVC_I$), allowing only one branch to continue the spike propagation. The selection is probabilistic due to intrinsic neuronal noise, which provides a source of stochasticity in the winner-take-all competition (Fig. 2b). In this model, syllable repetition is generated by connecting the syllable-chains to themselves at the branching points (Chang and Jin, 2009; Jin, 2009). In branched chain networks, the transitions between the syllable-chains are largely Markovian, and for repeating syllables this implies that repeat number distributions should be a decreasing function of the repeat number – in particular, the most probable (or "peak") repeat number will be one (Jin, 2009) (Fig.1c). However, many repeated syllables in Bengalese finch song have repeat distributions that are highly non-Markovian, with peak repeat numbers much larger than one (Fujimoto et al., 2011; Hampton et al., 2009; Jin and Kozhevnikov, 2011; Markowitz et al., 2013). This implies additional processes beyond synaptic chains contribute to generating non-Markovian repeated sequences.

Here we incorporate auditory feedback into the branching chain network model and show that, when this feedback is strong and adapting, non-Markovian repeat distributions emerge. In HVC, as in many sensory-motor systems, including the human speech system (Edwards et al., 2010; Wilson et al., 2004), the same neuronal populations that are responsible for the generation of the behavior also respond to the sensory consequences of that behavior, i.e. the bird's own song (BOS) (Lewicki and Konishi, 1995; Margoliash and Fortune, 1992; Prather et al., 2008; Sakata and Brainard, 2008). HVC receives much of its auditory input from NIf (Cardin et al., 2005; Coleman and Mooney, 2004; Fortune and Margoliash, 1995; Vates et al., 1996), which can provide real-time auditory feedback during singing (Fig. 2a) (Lei and Mooney, 2010). However, because of the time it takes to propagate motor commands to the periphery (30-50 ms) and process the subsequent auditory signals (15-20 ms) (Fig. 2a), auditory feedback is necessarily delayed relative to the motor activity that generated it (Wiener, 1948; Lashley, 1951; Sakata and Brainard, 2006, 2008; Bouchard and Brainard, 2013). This sensory-motor delay for HVC (45-70 ms) is on the order of the duration of a syllable, making it possible for auditory feedback to influence HVC motor programs for the transition dynamics between syllables (Hanuschkin et al., 2011; Sakata and Brainard, 2006, 2008) (Fig.2a).

The critical features of our framework for repeat generation are: (1) the population of neurons generating a repeated syllable receives a source of excitatory input in addition to the recurrent excitation from the sequencing network, and (2) the strength of this input adapts over time during repeat generation. We first tested the feasibility of this mechanism using biophysically detailed neural network models. To illustrate this model, we focus on generating sequences of the form 'ab$^n$c', where syllable 'a' transitions to syllable 'b', 'b' repeats a variable number of times (n), and transitions to 'c' (e.g. 'abbbbbbbc'). For concreteness, we model the adapting input as an auditory feedback signal to the network, though in principle this adapting input could reflect recurrent circuit-activity that is non-sensory. To incorporate auditory feedback into the previous model, each $HVC_{RA}$ neuron in chain-b is contacted by excitatory synapses carrying auditory inputs triggered by the production of syllable 'b' (Fig. 2c). We assume that the auditory synapses are made by axons from NIf, which is a major source of auditory inputs to HVC (Cardin et al., 2005; Coleman and Mooney, 2004; Fortune and Margoliash, 1995; Vates et al., 1996) and is selective to the bird's own song (Coleman and Mooney, 2004). When auditory feedback is present, the auditory synapses receive spikes from a Poisson process, assumed to be from the population of NIf neurons responding to syllable 'b' (Materials and Methods) (Fig. 2c). The auditory synapses are subject to short-term synaptic depression, resulting in gradual adaptation of responses to repeated inputs (Abbott et al., 1997; Markram and Tsodyks, 1996). Specifically, due to the synaptic depression, the average strength of the auditory inputs to chain-b decreases exponentially during the repeats of syllable 'b' (Materials and Methods).

In Figure 3, we show results from an example network in which the auditory input to chain-b is strong and the spiking dynamics produce repeats of syllable 'b' with large repeat numbers. A spike raster for a standard single run of the network is shown in Figure 3a. Once spiking was initiated in chain-a (through external current injection), spikes propagated through chain-a, and activated chain-b. Chain-b repeated a variable number of times before the spike activity exited to chain-c and stopped once it reached the end of chain-c. As chain-b continued to repeat, the synapses carrying the feedback signal weakened over time due to adaptation (Fig. 3b).

Analyzing multiple trials, we find that the probability of chain-b transitioning to itself (repeat probability) also decreases over time, though the repeat probability is only meaningful at the transition times – i.e. when the activity reaches the end of chain-b (Fig 3c). Examining the feedback strength at these transition times across the same trials allowed us to understand how the instantaneous feedback strength affects the repeat probability (Fig. 3d). Not surprisingly, we found that the repeat probability increases with the strengths of the auditory synapses. Repeat probability as a function of the feedback strength ($p_r$) could be well fit with the sigmoidal function (Fig. 3d, red curve)

$$p_r(A) = 1 - \frac{c}{(1 + \eta A^\nu)} \qquad (1)$$

where $A > 0$ represents the strength of the auditory synapses, $\eta, \nu > 0$ are parameters controlling the shape of the curve, and $0 < c < 1$ is a parameter for the repeat probability when there is no auditory feedback (i.e. $A = 0$), which is determined by the connection strengths of the network at the branching point. Note that, when the auditory input $A = 0$, the repeat probability is $p_r = 1 - c$, and conversely, as $A$ is large, $p_r$ approaches 1.

Initially, the strong auditory feedback biases the network toward repeating and so the repeat probability is close to 1. If the strong excitatory input resulting from auditory feedback were constant, the network would perseverate on repeating syllable 'b' indefinitely (a result of the positive feedback loop). However, because of the short-term synaptic depression, the auditory input to chain-b when syllable 'b' repeats decreases exponentially over time (Fig. 3b, red line; time-constant of $\tau = 148$ ms for this particular network). Even so, the repeat probability stays close to 1 as long as the auditory input is strong enough. Further weakening of the feedback reduces the repeat probability more significantly, making repeat-ending transitions to chain-c more likely. For this network, this process produced a repeat number distribution peaked at 6, as shown in Figure 3e. These results demonstrate that branched-chain networks receiving adapting excitatory inputs can generate repeat distributions that are non-Markovian.

**Statistical model for the repeat number distributions**

In our network model, the gradual reduction of excitatory drive from auditory feedback as a syllable is repeated reduces the probability that the syllable transitions to itself, and thus reduces the repeat probability. Eq. (1), describes the dependence of the repeat probability $p_r$ on the auditory input strength, $A$. The synaptic depression model tells us how $A$ changes with time. Sampling this at the transition times describes how $A$ changes with the repeat number, $n$. Putting this together, we can deduce (Materials and Methods) that

$$p_r(n) = 1 - \frac{c}{1+ab^n}, \tag{2}$$

where

$$a = \eta a_0^\nu$$

and

$$b = e^{-\nu T/\tau}.$$

Here $T$ is the syllable duration, $\tau$ is the time constant for the auditory adaptation, $a_0$ is the initial strength of the auditory feedback, and $\eta$ and $\nu$ are as in (1). Therefore, there are effectively three parameters (a, b and c) for how $p_r$ depends on $n$. We call Eq. (2) the *sigmoidal adaptation* model of repeat probability.

The network sequence dynamics can be represented with a state transition model, in which a single state corresponds to the repeating chain. The state can transition to itself with a probability $p_r(n)$ given by Eq. (2), or exit the state with probability $1 - p_r(n)$. This single state transition model can accurately fit the repeat number distributions generated by the network simulations with varying parameters as shown in Figure 4a (all fit errors below their respective benchmark errors, which characterize the fitting errors expected from the finiteness of the data set – see Materials and Methods).

This model contains the Markov model and a previously described 'geometric adaptation' model (Jin and Kozhevnikov, 2011) as special cases (Materials and Methods). Both of these models fail to fit the simulated data, even when a large number of states/parameters are used (Fig. 4b). On the other hand, we have shown that the sigmoidal model provides an accurate fit with a single state and a small number of parameters. Therefore, relative to other statistical models, the single-state transition model with sigmoidal adaptation parsimoniously and accurately replicates the syllable repetition statistics of our network model.

Using the single state transition model with sigmoidal adaptation, we explored how peak repeat numbers depend on the initial feedback strength and the adaptation strength (defined by the related parameter, $\alpha$, in the synaptic depression model, Materials and Methods) (Fig. 4c). Here we see that, for a given adaptation strength, there is a threshold feedback strength at which the peak repeat number is greater than 1, and this threshold increases with increasing adaptation strength. This demarcates the transition between Markovian (peak repeat number = 1) and non-Markovian (peak repeat number > 1) repeat distributions. Further increases in the feedback strength result in larger peak repeat numbers. Conversely, for a given feedback strength, increasing the adaptation strength results in a reduction of the peak repeat number. Together, these results demonstrate that a large range of peak repeat numbers can be generated through various combinations of feedback and adaptation strengths, and suggest that there is a threshold feedback strength required to produce non-Markovian repeat distributions.

**Sigmoidal adaptation model fits diverse repeat number distributions of Bengalese finch songs**

As demonstrated above, the single state transition model with sigmoidal adaptation (Eq. (2)) captures the essential features of our network model for repeating syllables, and fits well the repeat number distributions produced by the simulations. To examine whether these models can accurately describe syllable repeat number distributions of actual Bengalese finch songs, we recorded and analyzed the songs of 32 Bengalese finches. We identified the song syllables and obtained the syllable sequences (Materials and Methods). Our data set contains more than 82,000 instances of 281 unique syllables, of which 71 are repeating syllables. We fit the repeat number distributions for these syllables with the single state transition model with sigmoidal adaptation.

In Figure 5a, we show six examples of Bengalese finch repeat count histograms (grey bars) with different peak repeat counts (peak repeat count increases across plots i-vi.), and the best-fit model distributions (red lines). These examples show a range of distribution peaks and shapes, from small peak numbers with long rightward tails (i), to large peak numbers with tight, symmetric tails. Interestingly, we found that three repeated syllables had clear double-peaked distributions, with a prominent peak at repeat number 1 and another peak far away (two of which are displayed in panels ii and vi). We removed the single peak at repeat number 1 for these three syllables and only analyzed the longer repeat parts. The state transition with sigmoidal adaptation model does an excellent job of fitting the wide variety of peaks and shapes of the repeat distributions found in the Bengalese finches.

The results comparing the fit errors from the sigmoidal adaption model to benchmark errors across all 71 repeating syllables are shown in Figure 5b. The vast majority of fit errors from the feedback adaptation model are below their respective benchmark errors (86% of fit errors below the benchmark error), demonstrating that the model does an excellent job of fitting the diverse shapes of Bengalese finch song repeat number distributions. Therefore, the single state transition model with sigmoidal adaptation, and by extension the branched-chain model with adaptive auditory feedback, can successfully describe the syllable repeat number distributions in Bengalese finch songs.

**Removal of auditory feedback in Bengalese finches by deafening reduces peak repeat numbers**

In our framework, auditory feedback from the previous syllable arrives in HVC at a time appropriate to provide driving excitatory input to HVC neurons that generate the upcoming syllable. For repeated syllables, this creates a positive feedback loop which is responsible for generating peak repeat numbers greater than 1 (adaptation drives the process to extinction). Therefore, a key prediction is that without auditory-feedback driven excitatory input, the peak-repeat number should shift toward 1. To test this prediction, we deafened six Bengalese finches by bilateral removal of the cochlea, and analyzed the songs before and soon after they were deafened (2-4 days) (Materials and Methods).

We found that deafening greatly reduces the peak repeat-counts. For example, in Fig. 6a, we display spectrograms and rectified amplitude waveforms of the song from one bird prior to deafening (top) and soon after deafening (2-3 days post-deafening). We see that deafening reduces the number of times that the syllable (red-dashed box) is repeated. The acoustics of the syllables were not heavily degraded at this time, suggesting that the effects of deafening on syllable sequence can be separated from the effect on syllable phonology. The time course of repeat generation from this bird is examined in more detail in Fig. 6b, where we plot the median repeat counts per song of the syllable from Fig. 6a before deafening (black) and after deafening (red). Here we see that, even in the first songs recorded post-deafening, there is a marked decrease in the produced number of repeats. This data further exemplifies that repeat counts per song is generally stable across bouts of singing within a day both before and after deafening. Across days, repeat counts continued to slowly decline with time since deafening, though the co-occurrence of acoustic degradation of syllables makes these later effects difficult to interpret (Okanoya and Yamaguchi, 1997; Woolley and Rubel, 2002). Nonetheless, the rapidity of the effect of deafening underscores the acute function of auditory feedback in the generation of repeated syllables.

Similar results were seen across the other repeated syllables. Figure 6c shows the repeat number distributions for two additional birds before (black) and after (red) deafening. In these cases, deafening resulted in repeat number distributions that monotonically decayed, though the effect of deafening was larger for the repeat with larger initial repeat number (compare upper and lower panels of Fig. 6c). This suggests that degree to which deafening reduces peak repeat number depends on the initial repeat number. Across the 19 repeated syllables from 6 birds, deafening significantly reduced the number of consecutively produced repeated syllables (Fig. 6d, histogram on the right shows the difference in peak repeat number resulting from deafening, $p < 0.01$, sign-rank test, $N = 19$), although there was variability in the magnitude of the effect. We examined the change in peak repeat number resulting from deafening as a function of the peak repeat number before deafening (Fig. 6d, red dots correspond to data from individual syllables, overlapping points are horizontally offset for visual display). We found that the magnitude of decrease in peak repeat numbers after deafening grows progressively larger for syllables with greater peak repeat numbers before deafening ($R^2 = 0.81$, $p < 10^{-7}$, $N = 19$). This suggests that repeated syllables with larger repeat numbers are progressively more dependent upon auditory feedback for repeat production. Interestingly, after two days of hearing loss, one of the deafened Bengalese finches in our experiments had a repeat that was minimally affected by deafening, and several birds retained peaked repeat number around 2, not all the way to 1 as predicted for a Markov process. None-the-less, these deafening results are thus consistent with the hypothesis that the generation of repeated syllables is driven, in-part, by a positive-feedback loop caused by excitatory auditory input during singing.

**HVC auditory responses to repeated syllables gradually adapt**

A key prediction of the adaptive feedback model for repeat generation is that auditory responses of HVC neurons should decline over the course of repeated presentations of the same syllable. To test this hypothesis, we examined the properties of HVC auditory responses to repeated syllables in sedated birds (Materials and Methods). An example recording from a single HVC neuron in response to playback of the bird's own song (BOS) stimulus is presented in Figure 7a, which displays the stimulus spectrogram (top), the spike raster (middle), and the average spike rate in response to the stimulus (bottom). Two renditions of the repeated syllable 'b' are demarcated by red-dashed boxes, and we see that the evoked HVC auditory responses to repeated versions of the same syllable gradually declined.

The example presented above suggests that auditory responses to repeated presentations of the same syllable adapt over time. However, in the context of BOS stimuli, the natural variations that occur in syllable acoustics, inter-syllable gap timing, and in the identity of the preceding sequence, make it difficult to directly compare responses to different syllables in a repeated sequence. Therefore, to examine how responses to repeated syllables are affected by the length and identity of the preceding sequence, for each bird we constructed a stimulus set of long, pseudo-randomly ordered sequences of syllables (10,000 syllables in the stimulus, one prototype per unique syllable, median of all inter-syllable gaps used for each inter-syllable gap, derived from the corpus of each bird's songs, Materials and Methods). This stimulus allows a systematic investigation of how auditory responses to acoustically identical syllables depend on the length and syllabic composition of the preceding sequence (Bouchard and Brainard, 2013). Auditory responses at 18 multi-unit recordings sites in HVC from 6 birds were collected for this data set, which contained 40 unique syllables. Of these 40 syllables, 6 syllables in 4 birds (with 11 recording sites) were found to naturally repeat.

We used these stimuli to systematically examine how auditory responses to a repeated syllable depend on the number of preceding repeated syllables. We found that HVC auditory responses gradually declined to repeated presentations of the same syllable. In Figure 7b, for each uniquely repeated syllable (different syllables are colored from grey-to-red with increasing max repeat number), we plot the average normalized auditory response (mean ± s.e. across sites) to that syllable (e.g. 'b') as a function of the repeat number (e.g. repeat number 5 corresponds to the last 'b' in 'bbbbb'). Across HVC recordings sites and repeated syllables, the response to the last syllable declined as the number of preceding repeated syllables increased ($R^2 = 0.523$, $p < 10^{-10}$, N = 24, slope = -5%). Thus, auditory responses to repeated syllables gradually adapt as the number of preceding repeated syllables increases, providing confirmation of a key functional mechanism of the network model.

**Non-Markovian repeated syllables are loudest and evoke the largest HVC auditory responses**

To generate non-Markovian repeat distributions, we have proposed that the sequence generation circuitry is driven, in part, by auditory feedback that provides excitatory drive to sensory-motor neurons that control sequencing. Specifically, auditory feedback from the previous syllable arrives in HVC at a time appropriate to provide driving excitatory input to neurons that generate the upcoming syllable. This predicts that if HVC auditory responses are positively modulated by sound amplitude, feedback associated with louder syllables should provide stronger drive to the motor units, and thus generate longer strings of repeated syllables for a given rate of adaptation. This logic is supported by the sigmoidal adaptation model, which predicts a threshold auditory feedback strength at which the peak repeat number becomes greater than one (i.e. non-Markovian, Fig. 4b). Behaviorally, this predicts that non-Markovian sequences of repeated syllables should be composed of the loudest syllables in the bird's repertoire.

We tested this behavioral prediction by comparing the amplitudes of Bengalese finch vocalizations based on their repeat structure. Figure 8a plots the rectified amplitude waveforms (mean ± s.d.) of a few consecutively produced repetitions of a non-Markovian repeated syllable (black), a Markovian repeated syllable (red), and 'introductory' note (grey) from one bird. The non-Markovian repeated syllable is qualitatively louder than the other repeated vocalizations in the birds' repertoire. To quantitatively test this prediction, we measured the peak amplitude of the 281 unique syllables in our data set, and normalized this to the minimum peak amplitude across syllables (Materials and Methods). We categorized each syllable in our data set according to whether it was an introductory note (Intro), a non-repeated syllable (NR: repeats = 0), a Markovian repeated syllable (MR: peak repeat number = 1), or a non-Markovian repeated syllable (nMR: peak repeat number > 1). In Figure 8b, we plot the mean ± s.e. of the normalized peak amplitudes of these syllable groups across the data set. As exemplified by the data in Figure 8a, we found that non-Markovian repeated syllables were significantly louder than the other vocalizations in a bird's repertoire (***: $p < 10^{-3}$, **: $p < 10^{-2}$, sign-rank test, Bonferroni corrected for m = 3 comparisons). Therefore, syllables with non-Markovian repeat distributions are typically the loudest vocalizations produced by a bird.

If amplitude is a contributing factor to repeat generation, then HVC auditory responses should be positively modulated by syllable amplitude. However, previous work in the avian primary auditory system has found a population of neurons that is insensitive to sound intensity (Billimoria et al., 2008), and amplitude normalized auditory responses have been utilized in previous models of sequence encoding in HVC auditory responses (Drew and Abbott, 2003). Therefore, we next examined whether the increased amplitude of repeated syllables resulted in increased HVC auditory response to these syllables. We performed a paired comparison of normalized auditory responses to non-repeated syllables (NR) and non-Markovian repeated syllables (nMR) at the 11 sites where auditory responses to repeated syllables were collected (Fig. 8c). We found that repeated syllables had significantly larger auditory responses than non-repeated syllables ($p < 0.01$, sign-rank test, N = 11 sites). Thus, HVC auditory responses are sensitive to syllable amplitude, and repeated syllables elicit larger auditory responses than non-repeated syllables, likely due to being the loudest syllables that a bird sings. Therefore, the strong auditory feedback associated with these loud repeated syllables may be a key contributor to their non-Markovian repeat distributions.

**Discussion**

We have provided converging evidence that adapting auditory feedback directly contributes to the generation of repetitive vocal sequences in the Bengalese finch. A branching chain network model with adapting auditory feedback to the repeating syllable-chains produces repeat number distributions similar to those observed in the Bengalese finch songs. From the network model we derive the sigmoidal adaptation model for repeat probability, and show that it reproduces the repeat distributions of both the branching chain network and Bengalese finch data. Removal of auditory feedback by deafening reduced the peak repeat number, confirming one of the key features of the proposed mechanism. Furthermore, recordings in the Bengalese finch HVC show that auditory responses of HVC adapt to repeated presentations of the same syllable, providing evidence for another key feature of the proposed mechanism. Finally, we found that non-Markovian repeated syllables are louder than other syllables and elicit stronger auditory responses, suggesting that a threshold auditory feedback magnitude is required to generate long strings of repeated syllables, in agreement with modeling results. Together, these results implicate an adapting, positive auditory-feedback loop in the generation of repeated syllable sequences, and suggest that animals may directly use normal sensory-feedback signals to guide behavioral sequence generation with sensory adaptation preventing behaviorally deleterious perseveration.

In our framework, a positive feedback loop to a repeating syllable provides strong excitatory drive to that syllable and sustains high repeat probability. The strength of this feedback gradually reduces while the syllable repeats, preventing the network from perseverating on the repeated syllable. The combination of strong, positive feedback and gradual adaptation allows the production of non-Markovian repeat number distributions in the branching chain networks. We have conceptualized the adapting feedback as short-term synaptic depression of the NIf to HVC synapses resulting from auditory feedback. However, neither the exact source of the feedback nor the mechanism generating the adaptation is critical for our model. Indeed, the adaptation of auditory responses could arise from a variety of pre- and/or post-synaptic mechanisms anywhere in the auditory pathway, such as in the auditory forebrain (Beckers and Gahr, 2010), the auditory responses of NIf (Cardin et al., 2005; Coleman and Mooney, 2004; Fortune and Margoliash, 1995; Vates et al., 1996) or other auditory inputs to HVC such CM (caudal mesopallium) (Roy and Mooney, 2009), or within HVC itself. The biophysical origin of the auditory adaptation in HVC observed in our experiments remains to be determined.

Interestingly, after two days of hearing loss, one of the deafened Bengalese finches in our experiments maintained peaked repeat number distributions, and several birds retained peaked repeat numbers around 2, not all the way to 1 as predicted for a Markov process. One possible explanation is that there are several internal feedback loops to HVC within the song system that could contribute to repeating syllables. For example, there are direct anatomical projections from RA back to HVC (Roberts et al., 2008) as well as through the medial portion of MAN (mMAN) (Foster et al., 1997). Furthermore, there are connections from vocal brainstem nuclei to HVC through Uva and NIf (Schmidt, 2003; Wild, 1997). Although the signals transmitted through these internal feedback loops are poorly understood, they are likely to contribute to the temporal/sequential structure of song (Ashmore et al., 2005). These internal feedback loops may also contribute to, or even be the main routes of connecting the syllable encoding chains in HVC, rather than the direct connections between the chains within HVC as assumed in our network model. Furthermore, such internal feedback loops could be one site of adapting excitatory drive that contributes to the generation of non-Markovian repeats. However, our deafening results suggest that auditory feedback is a primary source of excitatory drive for repeat generation. Our modeling results should not change if such internal feedback loops are used instead of the direct connections for sequence generation, or instead of auditory feedback as the route of adapting positive feedback. Another possibility is the existence of multiple chains that produce syllables with similar acoustic features (Jin and Kozhevnikov, 2011). Such a "many-to-one mapping" could also explain the existence of residual non-Markovian features after deafening. Further understanding the role of these internal feedback loops and how they contribute to song generation is an important avenue for future research.

We observed that non-Markovian repeated syllables are typically the loudest syllables in a bird's repertoire. Furthermore, HVC responses to repeated syllables were significantly greater than responses to non-repeated syllables. Additionally, our modeling results demonstrate that non-Markovian repeat generation occurs only after a threshold auditory feedback strength has been reached. Together, these results suggest that syllable amplitude may be an important contributor to the generation of repeated syllables with non-Markovian distributions.

Our framework can be extended to allow auditory feedback to influence transition probabilities beyond repeated syllables. In general, because the auditory-motor delay in HVC due to neural processing is on the order of a syllable duration (Fig. 2a), auditory feedback from the previous syllable arrives in HVC at a time to contribute to the motor activity for the current syllable (Bouchard and Brainard, 2013; Sakata and Brainard, 2006, 2008). For a diverging transition of syllable 'a' to either 'b' or to 'c', as shown in Fig. 2b, auditory feedback from syllable 'a' can be applied to chain-b and chain-c. Depending on the amount of feedback on each chain, the transition probability to 'b' or 'c' can be enhanced or reduced by the feedback. Our model for repeating syllables (Fig. 2c) can be thought of as a special case of this general scenario, in which the repeating syllable-chain receives much stronger auditory input than the competing chain. The strong auditory feedback for repeated syllables may in part reflect synaptic weights that have been facilitated by Hebbian mechanisms operating on the repeated coincidence of auditory feedback with motor activity (Bouchard and Brainard, 2013). This framework is consistent with the observations that manipulating auditory feedback experimentally can change the transition probabilities (Sakata and Brainard, 2006). Auditory feedback plays a secondary role in determining the song syntax in our proposed mechanism. The allowed syllable transitions are encoded by the branching patterns of the chain networks. Auditory feedback biases the transition probabilities, to varying degrees for different syllable transitions. The secondary role of auditory feedback on the syntax could be the reason for the individual variations seen in a previous deafening experiment (Okanoya and Yamaguchi, 1997). Indeed, it was observed that one Bengalese finch maintained its song syntax 30 days after deafening (Okanoya and Yamaguchi, 1997). The secondary role of feedback in our model is in contrast to the model of Hanuschkin et al, who relied entirely on auditory feedback for determining syllable transitions (Hanuschkin et al., 2011). However, as in the Hanuschkin model, our model emphasizes the role of auditory feedback in shaping song syntax.

For motor control with continuous trajectories, such as reaching movements or articulation of single speech phonemes, it has been proposed that internal models estimate sensory consequences of motor commands, compare these estimates to actual sensory feedback, and use the difference as error signals to correct ongoing motor commands (Civier et al., 2010; Desmurget and Grafton, 2000; Houde and Nagarajan, 2011; Jordan, 1996). Along these lines, recent recordings in the auditory areas Field-L and CLM (caudolateral medopallium) of the zebra finch showed that, during singing, a subset of neurons exhibit activity that is similar to, but precedes, the activity induced by playback of the birds own song (Keller and Hahnloser, 2008). These data have led to the hypothesis that the songbird auditory system encodes a prediction of the expected auditory feedback ("forward model") used to cancel expected incoming auditory feedback signals (Jordan, 1996; Keller and Hahnloser, 2008; Lewandowski et al., 2013; Wolpert et al., 1995). According to such a forward model interpretation, as long as feedback matches expectation, auditory feedback does not reach HVC and therefore does not contribute to song generation during singing. At the surface, this seems at odds with our framework in which auditory feedback has a direct role in song generation, in particular for repeats. One possible resolution is that there is a forward model for the spectral content of syllables that is used to cancel the expected auditory feedback, but due to the increased loudness of non-Markovian repeated syllables, residual auditory input reaches HVC and contributes to song generation.

Some similarities between non-Markovian syllable repetitions in birdsong and sound/syllable repetitions in stuttered speech have been observed in the past (Helekar et al., 2003; Kent, 2000; Voss et al., 2010). In persons who stutter, repeating syllables within words ('to-to-to-today', for example) is a prominent type of speech disfluency (Boey et al., 2007; Van Riper, 1971; Zebrowski, 1994). Auditory feedback plays an important, but poorly understood, role in stuttered speech. For example, altering auditory feedback, including deafening (Van Riper, 1971), noise masking (Martin and Haroldson, 1979; Postma and Kolk, 1992), changing frequency (Stuart et al., 2008), and delaying auditory feedback reduces stuttering (Lane and Tranel, 1971). Conversely, delayed feedback can induce stuttering in people with normal speech (Lee, 1950; MacKay, 1986). Auditory processing may be abnormal both in zebra finches with abnormal syllable repetitions and in persons who stutter (Helekar et al., 2003). Our observation that deafening reduces syllable repetitions in Bengalese finch songs echoes the reduction of stuttering after deafening in persons who stutter (Van Riper, 1971). In general agreement with our proposed role of auditory feedback in repeat generation, some theories suggest that persons who stutter have weak feed-forward control and overly rely on auditory feedback for speech production (Civier et al., 2010). It will be interesting to see whether further quantitative analysis of the statistics of stuttered speech would reveal additional behavioral similarities, such as non-Markovian distributions and increased amplitude; to our knowledge no such examination exists. Such similarities could point to shared neural mechanisms with syllable repetition in birdsong, especially the possibility that auditory feedback plays a key role. However, our study also provides a cautionary note to the interpretation of repeated syllables in birdsong as 'stutters'. Our analysis shows that syllables with non-Markovian repeat distributions are loud and require strong auditory feedback. In contrast, syllables with Markovian repeat distributions are quieter and are less reliant on auditory feedback for their generation. We propose that it is the former type of syllable repetition that shares similarity to stuttering in humans.

## Materials and Methods

### The model neurons

The model neurons for the network simulations are a reproduction of those in previous works (Jin, 2009; Long et al., 2010). Below, we summarize the key aspects of these models. The reader is referred to these papers for exact details on the equations and constants. Since detailed information about the ion channels of HVC neurons is unavailable, we model both $HVC_{RA}$ and $HVC_I$ neurons as simple Hodgkin-Huxley type neurons, adding extra features to match available electrophysiological data. $HVC_I$ neurons exhibit prolonged tonic spiking during singing (Dutar et al., 1998; Kubota and Taniguchi, 1998). To simulate this we use a single-compartment model with the standard sodium-potassium mechanism for action potential generation along with an additional high-threshold potassium current that allows for rapid spike generation.

A distinctive feature of $HVC_{RA}$ neurons is that their activity comes in the form of precise bursts during song production (Hahnloser et al., 2002; Long et al., 2010). This bursting activity increases the robustness of signal propagation along chains of these neurons (Jin et al., 2007; Long et al., 2010). A recent study of the subthreshold dynamics of $HVC_{RA}$ neurons during singing suggests that this bursting is an intrinsic property of these cells (Long et al., 2010). We generate this intrinsic bursting behavior with a two-compartment model (Jin, 2009; Jin et al., 2007; Long et al., 2010). A dendritic compartment contains a calcium current as well as a calcium-gated potassium current. When driven above threshold, these currents produce a stereotyped calcium spike in the form of a sustained (roughly 5 ms) depolarization of the dendritic compartment. A somatic compartment contains the standard sodium-potassium currents for generating action potentials. These compartments are ohmically coupled so that a calcium spike in the dendrite drives a burst of spikes in the soma.

All compartments also contain excitatory and inhibitory synaptic currents. Action potentials obey kick-and-decay dynamics. All synaptic conductances start at 0. When an excitatory or inhibitory action potential is delivered to a compartment, the corresponding synaptic conductance is immediately augmented by an amount equal to the strength of the synapse. In between spikes, the synaptic conductances decay exponentially toward zero.

**Branching synfire chain model**

The network topology underlying all of the more advanced models below is the branching synfire chain network for HVC (Jin, 2009). $HVC_{RA}$ neurons are grouped into pools of 60 neurons. 20 pools are then sequentially ordered to form a chain. Except for the final pool, all neurons in a pool make an excitatory connection to every neuron in the next pool (Fig. 2b). The strengths of these synapses are randomly generated from a uniform random distribution between 0 and $G_{EEmax} = 0.09$ mS/cm$^2$. Because of this setup, activating the neurons in the first group sets off a chain reaction where each group activates the subsequent group, leading to a signal of neural activity propagating down the chain with a precise timing. There is one chain for every syllable in the repertoire of the bird. Activity flowing down a given chain drives production of the corresponding syllable through the precise temporal activation of different connections from $HVC_{RA}$ neurons to RA. To begin to impose a syntax on the song, the neurons in the final pool of one chain make connections to the initial pool of any chain whose syllable could follow its own. This branching pattern encodes the basic syllable transitions that are possible.

When the activity in an active chain reaches a branching point, all subsequent chains are activated, however only one should stay active – the syllable chosen next. This selection is achieved through lateral inhibition between the chains intermediated by $HVC_I$ neurons. There is a group of 1,000 $HVC_I$ neurons. Each $HVC_{RA}$ neuron has a chance of making an excitatory connection to each $HVC_I$ neuron with a probability $p_{EI} = 0.05$. Each of these connections has a strength randomly drawn from a uniform distribution between 0 and $G_{EImax} = 0.5$ mS/cm$^2$. In turn, each $HVC_I$ neuron has a chance of making an inhibitory connection to each $HVC_{RA}$ neuron with a probability $p_{IE} = 0.1$. The strengths of these connections randomly drawn from a uniform distribution between 0 and $G_{IEmax} = 0.7$ mS/cm$^2$. This setup gives a rough approximation of global inhibition on the $HVC_{RA}$ neurons which is what leads to the lateral inhibition between the chains that they comprise.

Noise is added to the network to make switching between chains a stochastic process. This noise is modeled as a Poisson process of spikes incident on each compartment of every neuron. The strength of each spike is randomly selected from a uniform distribution from 0 to $G_{noise}$ and every spike has an equal chance of being excitatory or inhibitory. Both compartments of $HVC_{RA}$ neurons receive noise at a frequency of 500 Hz; at the soma $G_{noise} = 0.045$ mS/cm$^2$, while at the dendrite $G_{noise} = 0.035$ mS/cm$^2$. The single compartment of the $HVC_I$ neurons receive noise at a frequency of 500 Hz with $G_{noise} = 0.45$ mS/cm$^2$. In $HVC_{RA}$ neurons, this leads to subthreshold membrane fluctuations of ~ 3 mV; in the HVCI neurons, the results is a baseline firing rate of ~10 Hz.

Each $HVC_{RA}$ neuron also receives an external drive that facilitates robust propagation of signals through the chains. This takes the form of a purely excitatory spike train modeled by a Poisson process with frequency 1000 Hz. The strength of each spike is chosen from a uniform random distribution from 0 to 0.05 mS/cm$^2$.

**Auditory feedback model**

We incorporate auditory feedback into the branching synfire chain model in a manner similar to the external drive used in (Jin, 2009) or the auditory feedback in (Hanuschkin et al., 2011). When a syllable is being produced and heard by the bird, some amount of auditory feedback can be delivered to any of the chains in the network in the form of external drives. The relative strength of this feedback drive between chains then biases transition probabilities so that auditory feedback plays an important role in determining song syntax.

The first piece in our model for auditory feedback is determining when auditory feedback from a specific syllable is active. We assume that the first few pools in every chain encode for the silence between syllables. Furthermore, once a syllable is being produced, there is a delay before auditory feedback begins that represents how long it takes for the bird to hear the syllable and process the auditory information. In our simulations, the activity of the 4th pool of every chain is monitored (by keeping track of the number of spikes in the previous 5 ms), with syllable production onset determined by when the population rate crosses a threshold of 43.3 Hz/neuron. After a delay of 40 ms, auditory feedback from that chain's syllable begins.

The auditory feedback takes the form of an external drive to all of the $HVC_{RA}$ neurons in a chain. Every chain can provide auditory feedback to every other chain, including itself. Thus, if there are N chains, then there are N$^2$ auditory feedback pathways. Denote the strength of the auditory feedback from chain i to chain j as $G_{ij}$. Every neuron in a chain will have N synapses, each one carrying the auditory feedback from one of the N chains in the network. The synapses carrying the auditory feedback from chain i to chain j have strengths drawn from a uniform random distribution between 0 and $G_{ij}$. Setting $G_{ij} = 0$ implies that there is no auditory feedback from chain i to chain j. When auditory feedback from a chain is active, the corresponding synapses are driven with Poisson processes at a frequency $f_{fdbk}$.

The model that each neuron receives only one synapse for each auditory feedback source is unrealistic. However, for computational simplicity, we model the feedback this way and consider each high-frequency synapse to be carrying spike trains from multiple sources. Since the kick-and-decay synapse model does not separate sources, this induces no real approximation. Auditory feedback parameters for Fig.2 were tuned to $f_{fdbk}$ = 1340 Hz and $G_{bb}$ = 1.9 mS/cm$^2$.

**Synaptic depression model**

To implement synaptic depression, we follow a phenomenological model put forward by Abbott et al. (Abbott et al., 1997). Whenever a synapse is used to transmit a spike, its strength $g$ is decreased by a constant fraction $\alpha$, so that $g \rightarrow (1 - \alpha)g$. The parameter $\alpha$ is referred to as the depression strength. In between spikes, the synaptic strength recovers toward its base strength $g_0$ with first order dynamics:

$$\tau_R \frac{dg}{dt} = -(g - g_0) \;.$$

The parameter $\tau_R$ is called the synaptic depression recovery time constant. If such a depressing synapse carries a spike train with a constant frequency $f$, the large-scale effect is an exponential decay to a steady-state strength where recovery and depression are balanced. The time constant of this decay as well as the steady-state strength can be expressed as functions of the model parameters: $\tau(\tau_R, \alpha, f)$ and $g_\infty(\tau_R, \alpha, f)$. See below for a derivation of the exact forms.

In our simulations with synaptic depression on the synapses carrying auditory feedback (in particular Fig.2), we use $\tau_R$ = 3.25 s and $\alpha$ = 0.006. It should be noted that, since these synapses actually represent the combined effect of multiple synapses (see above), these model parameters should not be taken as biologically representative. However, by matching the large-scale dynamics ($\tau$ and $g_\infty$) of the lower-frequency constituent synapses to that of the model synapse, one can find the more biologically relevant underlying depression parameters. Assume that each auditory feedback synapse represents the combined input of N constituent synapses, each carrying a spike train with a frequency $f/N$ so that the model synapse carries a spike train with frequency $f$. Matching the large-scale dynamics is then expressed as (primes representing biologically relevant parameters)

$$\tau(\tau_R', \alpha', f/N) = \tau(\tau_R, \alpha, f), \quad g_\infty(\tau_R', \alpha', f/N) = g_\infty(\tau_R, \alpha, f).$$

Since $\alpha$, $\tau_R$, and $f$ are known from the model, we can solve for $\alpha'$ and $\tau_R'$. With $N = 50$ this gives $\alpha' \approx 0.26$ and $\tau_R' \approx 3.75$ s – reasonable values for short-term depression in cortex (Abbott et al., 1997).

**Computational implementation**

Both the neural and synaptic depression models take the form of a large system of differential equations. A fourth-order Runge-Kutta scheme is used to numerically integrate these equations with custom code written in C++. When action potentials are generated during a time-step, synaptic conductances and synaptic depression dynamics are immediately updated before the next time-step is taken. All analysis is done with custom code in the MATLAB environment.

**Statistical model**

In branching chain networks, the transition probabilities at a branching point are determined by the relative strengths of the branching connections as well as the external inputs to the branching chains (Jin, 2009). The transition probability to a chain monotonically increases with the magnitude of excitatory input, saturating to 1 when the input is large (Fig. 3b). In our network model with short-term synaptic depression, the efficacy of the NIf drive to the repeating chain via auditory feedback decreases exponentially with time. After long synaptic adaptation, the NIf input reaches a residue equilibrium value that is much smaller than the initial value and can be ignored. Therefore, at the end of the n[th] repeat of the syllable, the NIf input reduces to

$$A(n) = a_0 e^{-nT/\tau}, \tag{M1}$$

where $a_0$ is the initial strength of the NIf input, $\tau$ is the time constant of the input decay, and $T$ is the duration of the syllable. Combining this with the dependence of the repeat probability on $A$, shown in Eq.(1), we find that the repeat probability after the n[th] repetition of the syllable is given by

$$p_r(n) = 1 - \frac{c}{1+\eta a_0^\nu e^{-n\nu T/\tau}} = 1 - \frac{c}{1+ab^n}, \tag{M2}$$

where

$$a = \eta a_0^\nu$$

and

$$b = e^{-\nu T/\tau}.$$

To systematically examine how the repeat number distribution depends on the strength $a_0$ of the auditory feedback and the adaptation strength $\alpha$, we used the sigmoidal adaptation model, Eq.(M1) and Eq.(M2), to generate repeat number distributions with combinations of these parameters. The decay time constant of the auditory feedback due to synaptic adaptation was set to

$$\tau = \frac{\tau_R}{1 - \tau_R f \log(1-\alpha)},$$

where $\tau_R$ is the recovery time constant and $f$ is the firing rate of NIf neurons during auditory feedback (see below). Besides $a_0$ and $\alpha$, all other parameters are set using those from the network simulations shown in Fig. 3 with T = 100 ms (approximately the length observed in the simulations). To simulate a repeat bout, we sequentially generate random numbers $x_k$ from a uniform distribution between 0 and 1 and compare each number to $p_r(k)$. The first time that $x_k > p_r(k)$ signifies that a further repeat does not occur, so the bout contains $k$ repeats. A distribution of repeats for a given $(a_0, \alpha)$ combination is produced by simulating the repeat bouts 10,000 times, and the results are shown in Fig. 4b, where we plot the peak repeat numbers for the distributions. Because the peak repeat number can go to infinity as adaptation strength goes to 0, for numerical stability we use a minimum adaptation strength of 0.001.

**Special cases of the sigmoidal adaptation model**

The sigmoidal adaptation contains two interesting special cases: (1) If we set the adaptation constant $\tau \to \infty$, which is equivalent to no adaptation of the auditory synapses, we have $b \to 1$ and the repeat probability becomes a constant, a hallmark of the Markov model for repeats. (2) If $c = 1$, which means the repeat probability is zero when the repeat number is large, and the initial auditory input is small such that when $ab \ll 1$, we have $p_r(n) \approx ab^n$, i.e. the repeat probability decreases by a constant factor with the repeat number. We call this the geometric adaptation of repeat probability. It was used to describe the repeating syllables in a previous work on the song syntax of the Bengalese finch (Jin and Kozhevnikov, 2011).

Any of these models can be extended to provide better fits to data by allowing multiple states. In these extended models, a repeated syllable is represented by multiple repeating states that all produce that syllable and are connected in series (Fig. 4b).

**Fitting repeat number distributions**

The probability of the syllable repeating $N$ times is given by

$$P(N) = (1 - p_r(N)) \prod_{n=1}^{N-1} p_r(n).$$

The observed repeat number probability $P_o(N)$ is computed by normalizing the histogram of the repeat numbers. The parameters $a$, $b$, $c$ are determined by minimizing the sum of the errors

$$\sum_N (P(N) - P_o(N))^2. \qquad (5)$$

while constraining the parameters ranges $0 < a < 10^8$, $0 < b < 1$, and $0 < c < 1$, using the nonlinear least square fitting function *lsqcurvefit* in MATLAB. To avoid local minimum in the search, 20 random sets of the initial values of the parameters were used for the minimization, and the best solution with the minimal square error was chosen.

**Comparing two probability distributions**

The difference between two probability distributions $p_1(n)$ and $p_2(n)$ is defined as

$$d = \frac{\max_n |p_1(n) - p_2(n)|}{\max_n (p_1(n), p_2(n))},$$

i.e. the maximum absolute differences between the two distributions normalized by the maximum of the two distributions (Jin and Kozhevnikov, 2011).

When fitting a functional form to a probability distribution, the difference between the empirical distribution and the fit is compared to a benchmark difference that represents the amount of error expected from the finiteness of data. For a given empirical distribution, the benchmark difference is computed by first randomly splitting into full data set into two groups of equal size and then computing the difference between the distributions resulting from each group. This process is repeated 1,000 times to produce a distribution of differences from simple resampling. The benchmark error is set at the 80[th] percentile of this bootstrapped distribution.

**Slow-scale depression dynamics**

Our model of synaptic depression characterizes the temporal dynamics of synaptic strength, $g$. Each synapse has a base strength, $g_0$. The depression model has two parameters: (1) depression strength, $\alpha$: fraction of strength lost at each spike; (2) recovery time constant, $\tau_R$: rate of exponential recovery toward $g_0$. Mathematically it can be described by two rules:

1. At a spike: $g \to (1 - \alpha)g$
2. Between spikes: $\tau_R \frac{dg}{dt} = -(g - g_0)$

We would like to characterize how this synapse will behave when transmitting a spike train that takes the form of a Poisson process. The analysis is simpler if we consider a regular spike train with frequency $f$ as an approximation. Fortunately, this should still give the average behavior for the Poisson process case. We begin by deriving an iterative map that takes the strength right before one spike and gives the strength right before the next.

Let the strength of the synapse right before a spike be $g$. Immediately after the spike, the strength will then be $(1 - \alpha)g$. Integrating the equation for recovery (from an initial condition $(t_i, g_i)$ to $(t_f, g_f)$) yields:

$$g_f = g_0 + (g_i - g_0)e^{-(t_f - t_i)/\tau_R} .$$

Since the spike train is assumed to be regular, we have $t_f - t_i = \frac{1}{f}$. And since the recovery starts from $g_i = (1 - \alpha)g$, the complete spike-to-spike iterative map is

$$g \to g_0 + ((1 - \alpha)g - g_0)e^{-1/\tau_R f} .$$

Using synaptic strength relative to $g_0$, i.e. $g = A g_o$ gives:

$$A \to 1 - e^{-1/\tau_R f} + (1 - \alpha)e^{-1/\tau_R f} A .$$

This iterative map has the form $A \to a + bA$, with $a = 1 - e^{-1/\tau_R f}$ and $b = (1 - \alpha)e^{-1/\tau_R f}$. If we start with $A = 1$, then this map has a closed-form solution:

$$A_n = \frac{a}{1-b} + \frac{(1-a-b)}{1-b} b^{n-1} .$$

This is a geometric decrease toward a steady-state value of $a/(1 - b)$ with a ratio of $r = b$. In terms of our model parameters, this is

$$g_\infty = g_0 \frac{1 - e^{-1/\tau_R f}}{1 - (1-\alpha)e^{-1/\tau_R f}}, \quad r = (1 - \alpha)e^{-1/\tau_R f} .$$

This discrete geometric decrease should be well-approximated by continuous exponential decay. The number of spikes needed to produce a fractional decrease of $e^{-1}$ is given by $r^n = e^{-1}$, so that $n = -\frac{1}{\log r}$. Since the inter-spike interval is $1/f$, the time constant of the continuous decay will thus be given by

$$\tau = \frac{n}{f} = \frac{1}{f \log[(1-\alpha)e^{-1/\tau_R f}]} = \frac{\tau_R}{1-\tau_R f \log(1-\alpha)}.$$

While this derivation is for a regular spike train, simulations (not shown) verify that it is also fits the large-scale dynamics of a Poisson spike train with the same mean frequency.

**Animals**

32 birds were used in this study. All 32 birds contributed to the behavioral analysis (Fig.5). Of these 32 birds, six birds were used in the deafening studies. A different subset of six birds were used in the electrophysiology experiments. During the experiments, birds were housed individually in sound-attenuating chambers (Acoustic Systems, Austin, TX), and food and water were provided ad libitum. 14:10 light:dark photo-cycles were maintained during development and throughout all experiments. Birds were raised with a single tutor. All procedures were performed in accordance with established animal care protocols approved by the University of California, San Francisco Institutional Animal Care and Use Committee (IACUC).

**Song collection and analysis**

All behavioral analyses, as well as stimulus creation, were done using custom code written in Matlab (The Mathworks, Natick, MA). Individual adult male Bengalese finchs were placed in a sound-attenuating chamber (Acoustic Systems, Austin, Tx) to collect audio recordings. An automated triggering procedure was used to record and digitize (44,100 Hz) several hours of the bird's singing. These recordings were then scanned to ensure that more than 50 bouts were obtained. Bouts were defined as continuous periods of singing separated by at least 2 seconds of silence. Bengalese finch songs typically consist of between 5-12 distinct acoustic events, termed syllables, organized into probabilistic sequences. Each bout of singing consists of several renditions of sequences, with each sequence containing between 1 and approximately 40 examples of a particular syllable. The syllables from 15-50 bouts were hand labeled for subsequent analysis.

**Deafening**

Birds were deafened by bilateral cochlear removal (Brainard and Doupe, 2001; Konishi, 1965). Complete removal of the cochlea, including the distal end of the auditory nerve, was visually confirmed using a dissecting microscope. After cochlear removal, some birds showed signs of vestibular disturbance that usually resolved in the first few days after surgery. Extra care was taken to ensure that such birds had easy access to seed and maintained full crops. Birds did not exhibit difficulty in perching, feeding, or interacting with other birds after returning to their home cages.

**Electrophysiology**

The electrophysiological results presented in this study were collected as part of a larger study investigating how sequences and syllable features are encoded in HVC auditory responses. The data used in this study and the associated methods have been described previously (Bouchard and Brainard, 2013). Briefly, for neural recordings, birds were placed in a large sound-attenuating chamber (Acoustic Systems, Austin, TX) and stereotaxically fixed via a previously implanted pin. During electrophysiological recordings, birds were sedated by titrating various concentrations of isoflurane in $O_2$ using a non-rebreathing anesthesia machine (VetEquip, Pleasanton, CA). Throughout the experiment, the state of the bird was gauged by visually monitoring the eyes and respiration rate using an IR camera. Sites within HVC were at least 100 μm apart and were identified based on stereotaxic coordinates, baseline neural activity, and auditory response properties. Experiments were controlled and neural data were amplified with an AM Systems amplifier (x1000), filtered (300-10,000 Hz), and digitized at 32,000 Hz.

**Playback of auditory stimuli**

Stimuli were band-pass filtered between 300-8,000 Hz and normalized such that BOS playback through a speaker placed 90 cm from the head had an average sound pressure level of 80 dB at the head (A scale). Each stimulus was preceded and followed by 0.5-1 s of silence and a cosine modulated ramp was used to transition from silence to sounds. The power spectrum varied less than 5 dB across 300-8,000 Hz for white-noise stimuli. All stimuli were presented pseudo-randomly.

**Creation of pseudo-random stimuli**

To probe how repeated syllables are encoded in the population of HVC neurons, we used a stimulus set that consisted of 10 strings of 1000 pseudo-randomly ordered syllables was constructed. The details of this stimulus are described previously (Bouchard and Brainard, 2013). Briefly, for each bird, natural sequences (i.e. sequences produced by a given bird) and non-natural sequences (i.e. sequences that were never produced by a bird) of length 1 through 10 were concatenated with equal probability into 10 strings of 1000 syllables. For each syllable in the birds repertoire occurring in these stimuli, a single 'prototype' syllable was used based on the distributions of acoustic features of that syllable. The median of all inter-syllable gaps was used for each gap. BOS stimuli created with these elements (synthesized BOS, prototype syllables and median gaps) elicit HVC auditory responses of comparable magnitude to normal BOS stimuli. Additionally, responses to single syllables preceded by the same long sequences in the pseudo-random stimuli are not significantly different from responses in synthesized BOS. Thus, these stimuli isolate sequence variability from other sources of variability in song, and allow investigating how HVC auditory responses to individual syllables are modulated by the preceding sequence.

**Spike sorting, calculation of instantaneous firing rates, and responses to individual syllables**

Single units were identified events exceeding 6 standard deviations from the mean and/or were spike sorted using in house software based on a Bayesian inference algorithm. Multi-unit neural data were thresholded to detect spikes more than 3 standard deviations away from the mean. Both single and multi-unit spike times were binned into 5 ms compartments and then smoothed using a truncated Gaussian kernel with a standard deviation of 2.5 ms and total width of 5 ms.

To characterize the responses to individual target syllables, we defined a response window, which started 15 ms after the onset of the syllable and extended 15 ms after the offset of that same syllable.

**Statistics**

All statistical tests were performed using either paired sign-rank tests or unpaired rank-sum tests. Throughout the paper, results were considered significant if the probability of Type I errors was $\alpha < 0.05$. Bonferroni corrections were used to adjust $\alpha$-values when multiple comparisons were performed.

**Acknowledgements**

This research was supported by NSF Grant No. IOS-0827731 and the Huck Institute of Life Sciences at the Pennsylvania State university (DZJ) and NIH R01 DC006636, NSF IOS-0951348 and the Howard Hughes Medical Institute (MSB).

**Author Contributions**

Computational model: JW, DZJ. Experiments: KEB, MSB. Data analysis: KEB. Wrote paper: DZJ, KEB, JW, MSB.

**Figure Legends**

**Figure 1. Bengalese finch song and the generation of sequences**

a. Diagram of sensory-motor circuit for sequence generation. An internal motor program generates transitions between actions ('a', 'b', 'c', etc) while sensory feedback from the actions (motor outputs) impinges on the motor program.  b. Example of Bengalese finch song. Spectrogram (power at frequency vs. time) of an adult Bengalese finch song, which consists of several syllables (denoted with letters) produced in probabilistic sequences. A prominent feature of Bengalese finch songs is the presence of syllable repetitions, some with long repeat sequences (e.g. syllable 'b').  c. Probability distribution of repeat counts for syllable 'b' from an individual Bengalese finch (red), and the predicted probability distribution for a Markov process using first order transition probabilities.

**Figure 2. Avian song system and branched chain network with adapting auditory feedback.**
a. Diagram of the avian song system. Nucleus HVC (used as proper name) is a sensory-motor integration area that receives auditory input from high-level auditory nuclei such as NIf (nucleus interfacealus), and sends temporally precise motor controls signals to nucleus RA (robust nucleus of the arcopallium), which projects to the vocal brainstem areas. There is a pre-motor latency of 30-50 ms (ΔT Motor) between activity in HVC and subsequent vocalization. Additionally, there is a latency of 15-20 ms (ΔT Auditory) for auditory activity to reach HVC.

This makes for a total auditory-motor latency between pre-motor activity and resulting auditory feedback of 45-70 ms. b. Example of a branch point in a probabilistic sequence (left). Syllable 'a' can transition to either syllable 'b' or 'c'. Such probabilistic sequences can be produced by a branched chain network (right). Here, each syllable is produced by a syllable-chain, in which groups of $HVC_{RA}$ neurons (grey dots in red ovals, grouped in grey rectangles for a given syllable) are connected unidirectionally in a feed-forward chain (black lines with triangles are excitatory connections). The end of chain-a connects to the beginning of chain-b and chain-c. Spike activity propagates through chain-a and drives downstream neurons in RA to produce syllable a. At the end of chain-a, the activity continues to chain-b or chain-c via the branched connections. Only one syllable chain can be active at a time, as enforced by winner-take-all mechanisms mediated through local feedback inhibition from the $HVC_I$ neurons (red lines are inhibitory connections). c. The branched chain network with adapting auditory feedback for generating repeating sequences of syllable b. The end of chain-b reconnects to the beginning of itself and chain-c. Auditory feedback from syllable 'b' is applied to chain-b, and biases the repeat probability when the activity propagates to the branching point. The feedback is weakened as syllable 'b' repeats due to use dependent synaptic depression.

**Figure 3. Strong, adapting auditory feedback produces peaked repeat distributions in branching chain neural networks.**

a. Raster plot of the spikes of the neurons in the network model. Neurons are ordered according their positions in the chains. Each dot represents the spike time of a neuron. Spikes are subsampled and image smoothed so that darker areas represent stronger activity at a particular location/time. Spikes propagate from chain-a to chain-b. Chain-b repeats a variable number of times before activity exits to chain-c. b. The strength of auditory synapses decreases once they are activated by feedback auditory input. The red line is the fit to an exponential function with a decay time constant $\tau = 148$ ms. There is auditory feedback from syllable-b to $HVC_{RA}$ neurons in chain-b during the times indicated by white areas. c. The probability of chain-b repeating decreases as the repeat number increases. The probabilities are computed over 100 simulations. d. The repeat probability of chain-b as a function of the average synaptic strength of the auditory inputs that the chain receives at the transition time. The bars are standard deviations. The red line is a fit with a sigmoidal function. e. The probability distribution of the repeat numbers of syllable 'b' – probabilities computed over 1,000 simulations.

**Figure 4. Sigmoidal adaptation model of repeats and model predictions.**
a. Six example repeat count histograms (black bars) from the neural network simulations with adapting auditory feedback and the best fit distributions from the sigmoidal adaptation model (red lines). The decay constants of the auditory feedback and the syllable lengths are varied to produce different repeat number distributions. Syllable lengths are changed by altering the number of groups per chain. All fit errors are smaller than benchmark errors. b. Best fit geometric adaptation models for the first histogram in a. With geometric adaptation, the probably of a state repeating is decreased by a constant factor with each consecutive repeat (Materials and Methods): (i) single state; (ii) two-states, both repeating; (iii) multiple-states, only final state repeating. In all cases, numbers on arrows are initial transition probabilities while the number in parenthesis is the constant adaptation factor. c. Comparison of model fits. Red is the best fit of the sigmoidal adaptation model with one state. Other colors are best fits of the corresponding models in b. The sigmoidal adaptation model provides a superior fit while only requiring a single state. d. Peak repeat number as a function of the initial auditory feedback strength and the adaptation strength generated using the sigmoidal adaptation model. The peak repeat number increases for increasing initial feedback strength and decreases for increasing adaptation strength. For a given adaptation strength, there is a threshold feedback strength at which repeat distributions become non-Markovian (i.e. peak repeat number >1).

**Figure 5. Sigmoidal adaptation model fits diverse repeat number distributions of Bengalese finch songs**
a. Six example Bengalese finch repeat count histograms (grey bars) and the best-fit model distributions (red lines). Peak repeat count increases from left-to-right and down columns. Distribution marked with (*) provide two examples of repeat distributions that have clear double peaks. For these cases, the peaks at repeat number 1 are excluded. b. Scatter plot of fit error vs. benchmark error. Each red circle corresponds to the distribution for one repeated syllable from the song database. The fit errors are smaller than the benchmark errors in the vast majority of cases (86%).

**Figure 6. Removal of auditory feedback in Bengalese finches by deafening reduces peak repeat counts**

a. Example spectrograms and rectified amplitude waveforms (blue traces) for the song of one bird before (top) and after (bottom) deafening. Red dashed boxes demarcate the repeated syllables. b. Median repeat counts per song of the syllable from before deafening (black) and after deafening (red). Rotated probability distributions at the right hand side display the repeat counts across all recorded songs before (black) and after (red) deafening. c. Additional examples of repeat distributions pre- (black) and post- (red) deafening. For syllables that were repeated many times, deafening caused sharp reductions in repetitions, resulting in repeat number distributions that are more Markovian (upper graphs). Deafening had less of an effect on syllables that were repeated fewer times (lower graphs). d. Peak repeat numbers before deafening vs. the differences in peak repeat numbers before and after deafening. Red dots correspond to syllables and black line is from linear regression. Rotated histogram on the right shows the distribution of the differences. Deafening results in a significant decrease in the peak repeat numbers (Wilcoxon sign-rank test, $p < 10^{-2}$, $N = 19$), with larger decreases in peak repeat numbers for syllables that were repeated many times before deafening ($R^2 = 0.81$, $p < 10^{-7}$, $N = 19$).

**Figure 7. HVC auditory responses to repeated syllables gradually adapt**

a. Example auditory recording from a single unit in HVC in response to playback of the BOS (bird's own song) stimulus. Top panel is the song spectrogram with labeled syllables. Middle panel is the spike raster from 45 trials. Bottom plot is the average response rate across trials. Adaptation of HVC auditory responses to a repeated syllable (demarcated by red-dashed lines) is observed. b. Responses to the last syllable in a repeat as a function of the repeat number. Data are presented as mean ± s.e. of normalized auditory responses across sites for a given repeated syllable (11 sites in 4 birds, 6 repeated syllables). Data are colored from grey-to-red with increasing peak repeat number. Across syllables and sites, the response to the last syllable in a repeat declines with increasing repeat number. Black line is from linear regression ($R^2 = 0.523$, $p < 10^{-10}$, $N = 24$, slope = -5%).

**Figure 8. Non-Markovian repeated syllables are loudest and evoke the largest HVC auditory responses**

a. Mean ± s.d. amplitude waveforms for a non-Markovian repeated syllable (black), a Markov repeated syllable (red), and an intro note (grey) from the songs of one bird. b. Mean ± s.e. normalized peak amplitudes of song vocal elements. Intro notes (Intro), non-repeated syllables (NR), Markov-repeated syllables (MR, peak repeat number = 1), and non-Markovian repeated syllables (nMR, peak repeat number > 1). non-Markovian repeated syllables are significantly louder than other vocalizations ($p < 10^{-3}$, $p < 0.01$, Wilcoxon sign-rank test, Bonferroni corrected for m = 3 comparisons). c. Paired comparison of normalized auditory responses to non-repeated syllables (NR) and non-Markovian repeated syllables (nMR). Repeated syllables illicit larger auditory responses. ($p < 0.01$, Wilcoxon sign-rank test, N = 11 sites). Circles: data; square: median.

Figure 1.

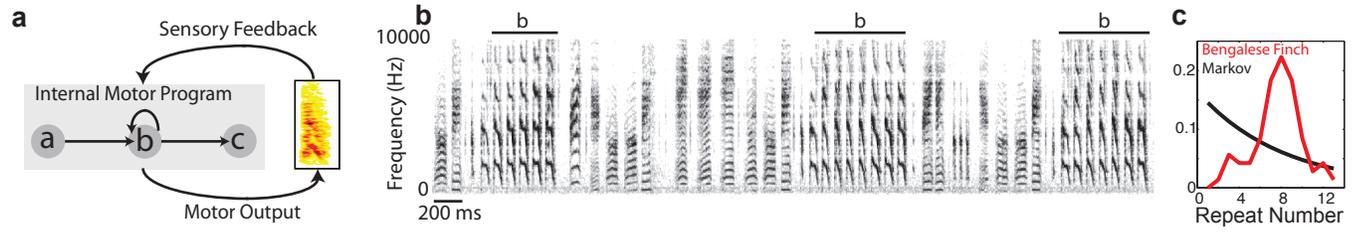

Figure 2.

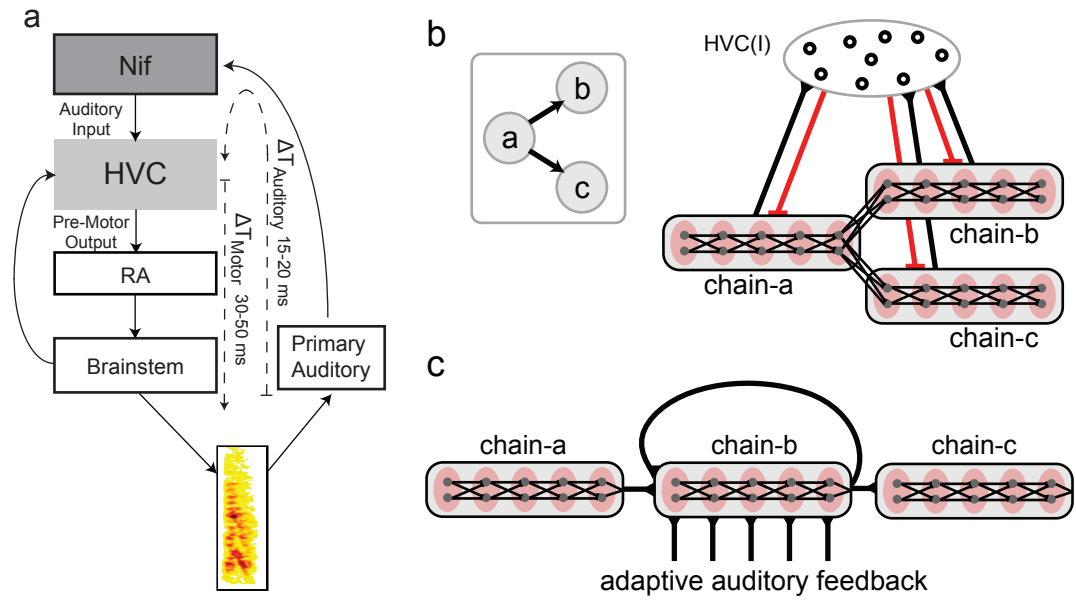

Figure 3.

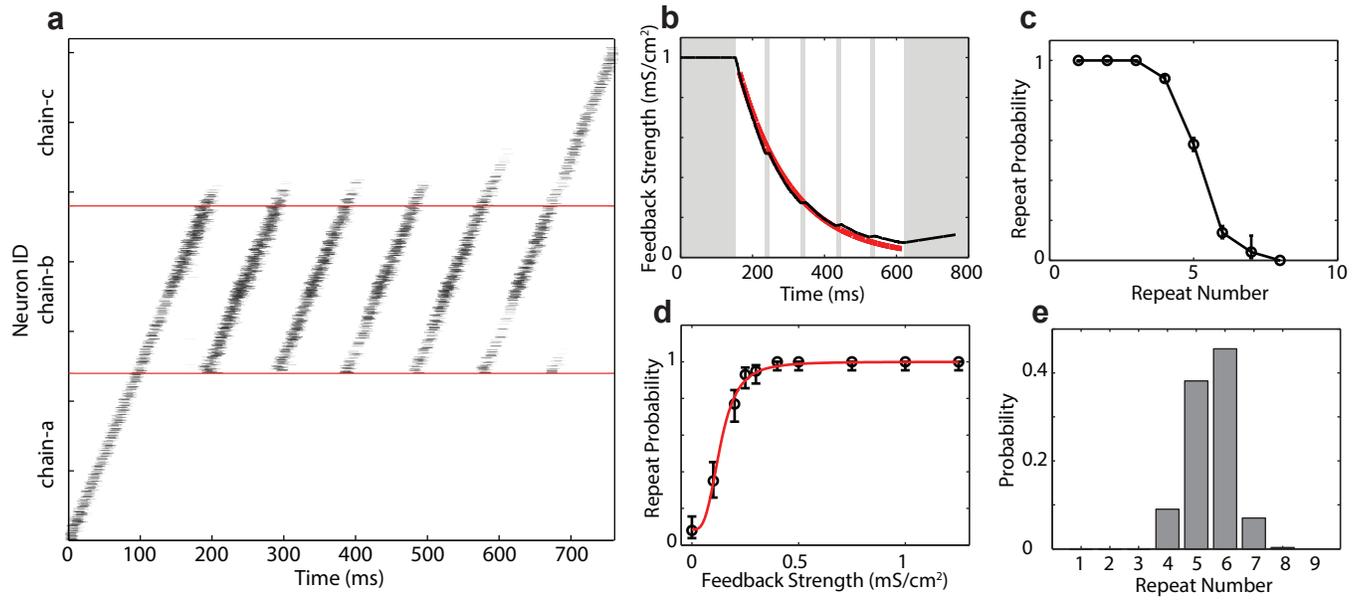

Figure 4.

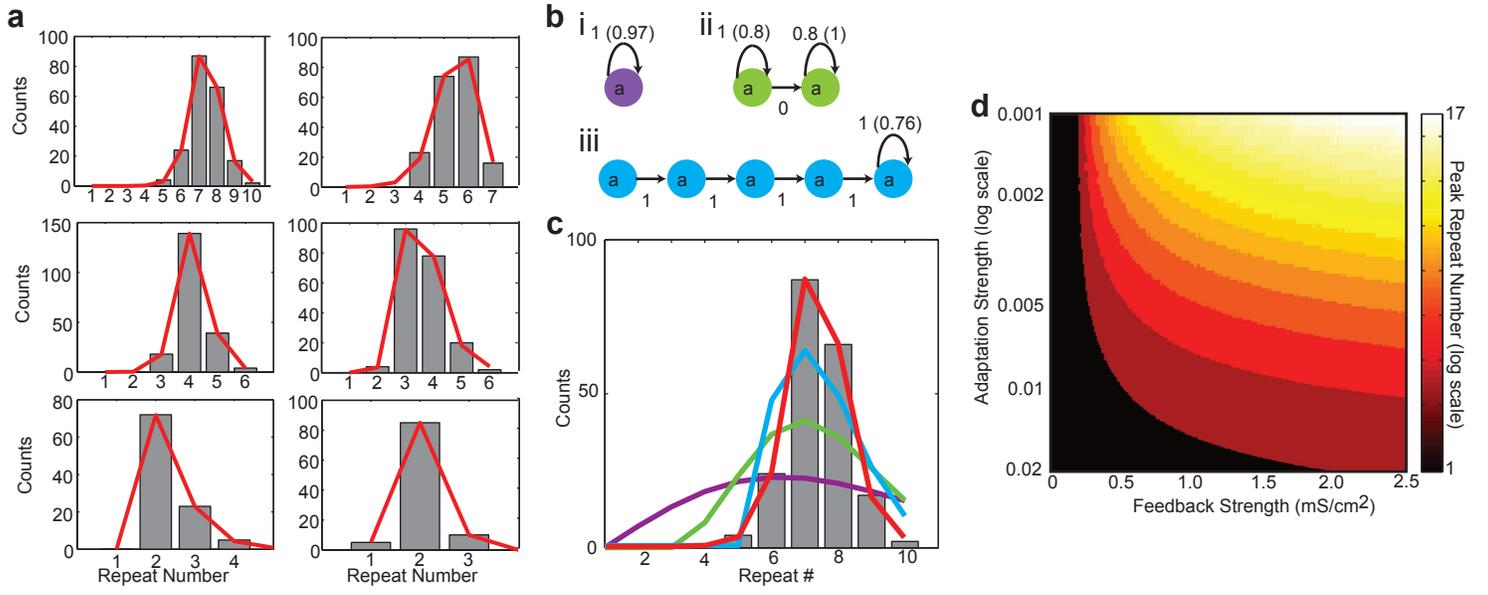

Figure 5.

a 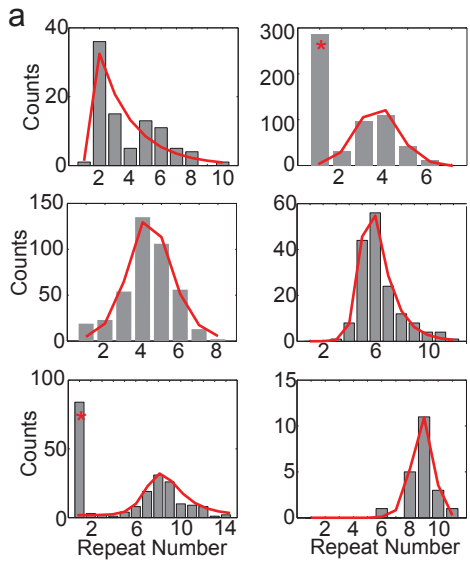
b 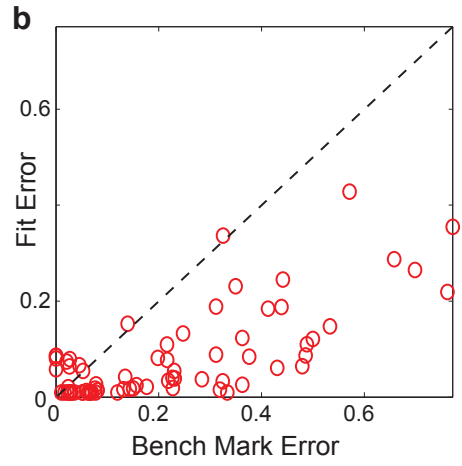

Figure 6.

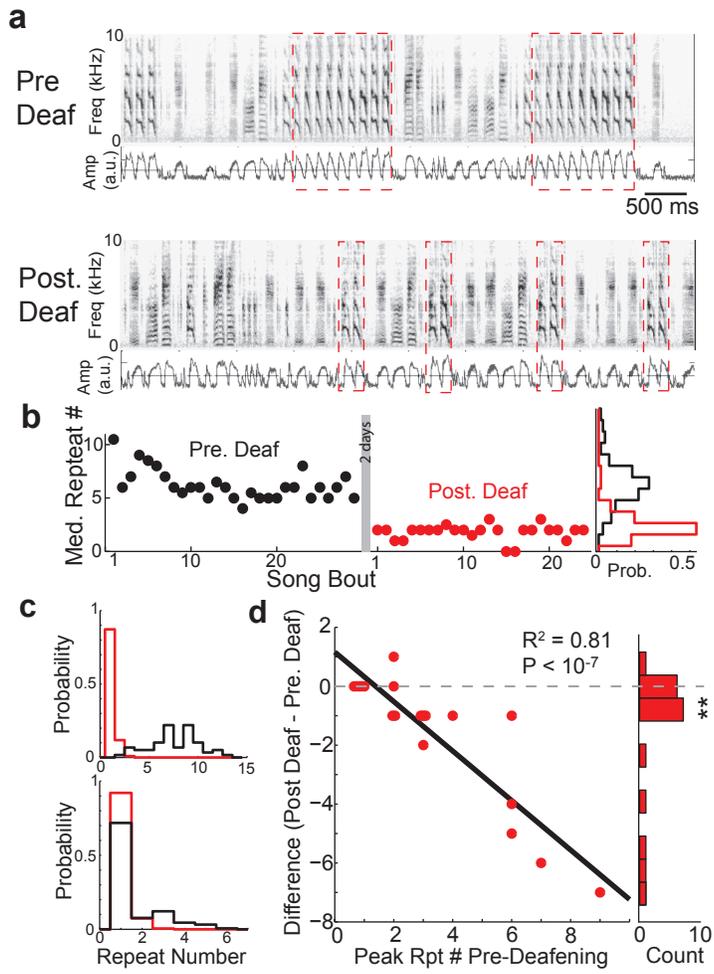

Figure 7.

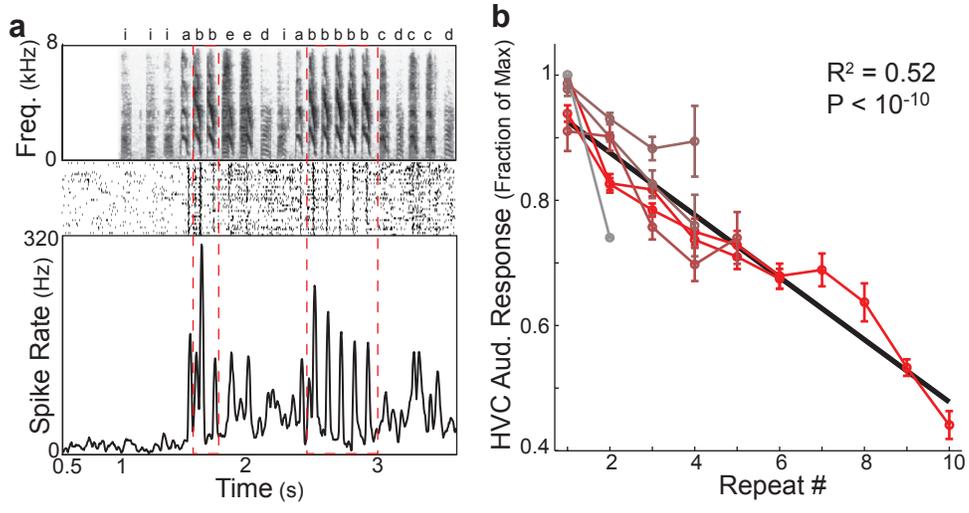

Figure 8.

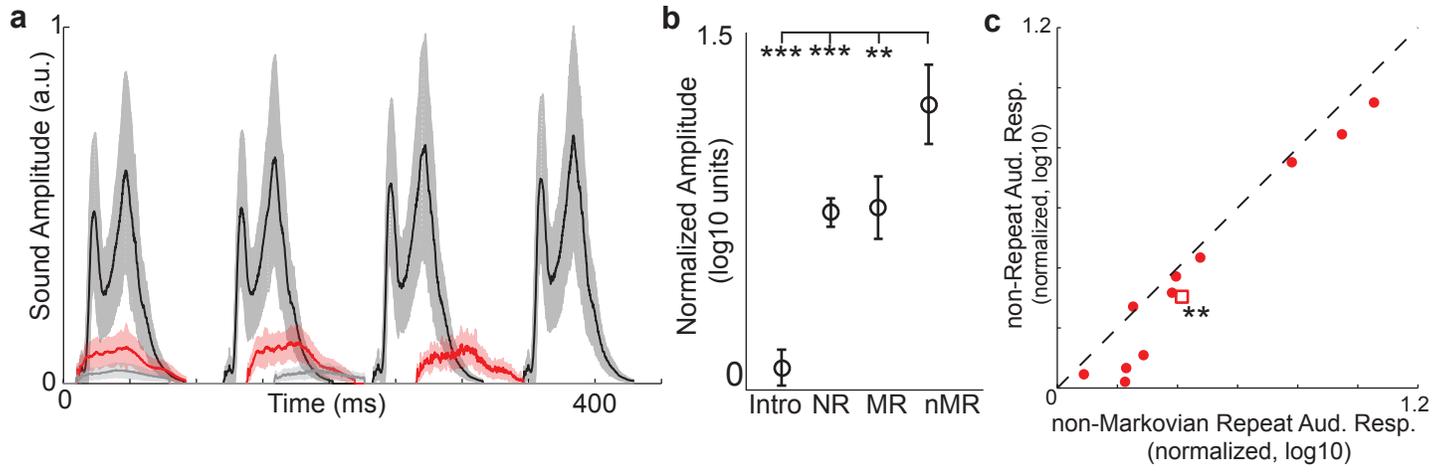